\documentclass[12pt]{article} 

\usepackage{latexsym,amsmath,amssymb,theorem,epsfig}

\DeclareFontFamily{OT1}{rsfs10}{}
\DeclareFontShape{OT1}{rsfs10}{m}{n}{ <-> rsfs10 }{}
\DeclareMathAlphabet{\mathscript}{OT1}{rsfs10}{m}{n}

\numberwithin{equation}{section}

\newcommand{\ns}{\normalsize}

\def\l{\lambda}

\theoremstyle{plain}

{\theorembodyfont{\rmfamily} }

\begin{document}

%%%%%%%%%%%%%%%%%%%%%%%%%%%%%%%%%%%%%%%%%%%%%%%%%%%%%%%%%%%%%%%%%%%%%%

\begin{titlepage}

\vspace{-5cm}

\title{
  \hfill{\ns }  \\[1em]
   {\LARGE Stable and Metastable Cosmic Strings in Heterotic M-Theory}
\\[1em] }
\author{
   Evgeny I. Buchbinder
     \\[0.5em]
   {\ns School of Natural Sciences, Institute for Advanced Study} \\[-0.4cm]
{\ns Einstein Drive, Princeton, NJ 08540}\\[0.3cm]}

\date{}

\maketitle

\begin{abstract}

We address the question of finding stable and metastable cosmic strings
in quasi-realistic heterotic M-theory compactifications
with stabilized moduli and de Sitter vacua. 
According to Polchinski's conjecture, the only
stable strings in the absence of massless fields are Aharonov-Bohm strings. 
Such strings could potentially be created in heterotic compactifications
as bound states of open membranes, five-branes wrapped on four-cycles
and solitonic strings. However, in generic compactifications,
the process of moduli stabilization can conflict production of Aharanov-Bohm strings.
In this case, 
heterotic cosmic strings will have to be unstable under breakage on monopoles. 
We estimate the monopole masses and find that they are big enough
so that the strings can be metastable with a sufficiently long lifetime.
On the other hand, if we allow one or more axions to remain 
massless at low energies, stable global strings can be produced.

\end{abstract}

\thispagestyle{empty}

\end{titlepage}

%%%%%%%%%%%%%%%%%%%%%%%%%%%%%%%%%%%%%%%%%%%%%%%%%%%%%%%%%%%%%%%%%%%%%%%%%%%%%%%%%%%%%%%%%%%%%%%%%%%%%%%%

\section{Introduction}

%%%%%%%%%%%%%%%%%%%%%%%%%%%%%%%%%%%%%%%%%%%%%%%%%%%%%%%%%%%%%%%%%%%%%%%%%%%%%%%%%%%%%%%%%%%%%%%%%%%%%%%%%

In~\cite{Polchinski}, Polchinski proposed a classification of various
cosmic strings in the context of superstring theory and conjectured 
that any theoretically possible decay should be allowed. 
This conjecture implies that in the framework of string theory
there are only two types of absolutely stable cosmic strings. 
The first type is global strings which are charged under a massless axion.
The second type is Aharonov-Bohm strings~\cite{AB}. These strings can exist only 
if there is a particle that picks up a fractional phase around the string
and does not couple to the low-energy gauge group. For example, 
in string compactification models with all moduli, including all axions, stabilized,
Aharonov-Bohm strings, provided they exist, are the only stable strings.
To find a concrete decay mechanism of other types of strings
in a given model can be very complicated.
Nevertheless, if Polchinski's conjecture is correct it always exists.
For example, in some heterotic $SO(32)$ compactifications, a fundamental 
string seems stable though it is neither global nor Aharonov-Bohm. 
However, in~\cite{Polchinski} it was demonstrated the existence 
of open heterotic $SO(32)$ strings which in four dimensions end on monopoles. 
This provides a concrete mechanism how a fundamental $SO(32)$ string 
can break on monopoles. If this conjecture is correct, it is
likely that cosmic strings in compactification and brane models 
are only metastable. A systematic analysis of of fundamental 
and Dirichlet cosmic strings was performed in~\cite{CMP}.
It was shown that in type IIB models with many Klebanov-Strassler~\cite{KS} throats 
it is possible to have brane/anti-brane inflation~\cite{KKLMMT} followed by 
de Sitter (dS) vacua~\cite{KKLT}
and production of cosmic F- and D-strings. In fact, the strings produced 
this way are only metastable. Various properties of these strings were 
studied in~\cite{Dvali1, Dvali2, Dvali3, GHK}.

In this paper, we attempt to describe how stable and metastable cosmic 
strings can arise in heterotic M-theory~\cite{HW1, HW2, Wittenstrong}. 
Such compactifications are
very attractive from the phenomenological viewpoint. 
Recent results~\cite{Volker1, Volker2, Donagi, Volker3} 
show the possibility of obtaining the Standard Model spectrum with no extra exotic
matter in this framework. Moduli stabilization and inflationary properties
of heterotic M-theory were studied in~\cite{CK, BO, BCK, Raise, Myinflation, Theirinflation, Vaman, Ward}. 
Cosmic strings in these models were studied in~\cite{Cosmic, Beckercosmic}.
The natural cosmic string candidates are open membranes, five-branes wrapped on four-cycles
in compactification space and solitonic strings as well as their bounds states.
All these strings have two major universal sources of instability. The first one is that 
they bound axionic domain walls. If the axion receives a mass
the strings will rapidly collapse~\cite{Vilenkin}. The
second instability is the possible breakage on monopoles. The latter instability is not as severe
as the former. If the square of the mass of the monopoles is much bigger than the 
string tension, the string will be metastable with a long lifetime. 

To understand exactly which string or bound states can arise in a given model 
it is important to understand the potential for all the axions. Such potential 
always arise in stabilizing moduli. This is why it seems hard to ask the question 
about stability of various strings outside of the context of moduli stabilization. 
In sections 2 and 3, we discuss moduli stabilization in heterotic M-theory.
The reason is twofold. First, as just explained, it is important 
for understanding what kind of axionic potentials appear in the known methods
of moduli stabilization. Second, the tension of potential cosmic string candidates 
is moduli dependent. Therefore, it is important to set-up moduli stabilization
procedure to be consistent with having the tension of possible strings
within the observational bound. In~\cite{Cosmic}, it was shown that it possible to stabilize a five-brane
close to the visible sector so that an open membrane stretched between
this five-brane and the visible brane will appear as a string with a small tension
In section 2, we show that it is possible to stabilize the Calabi-Yau 
Kahler moduli in such a way that one or more two-cycles are much smaller
than the Calabi-Yau scale.
This implies the possibility 
that a five-brane wrapped on a four-cycle will also appear as a string with a small 
tension. In section 3, we discuss the volume stabilization, 
Fayet-Iliopoulos (FI) terms 
and dS vacua. In section 4, we discuss stability of cosmic strings. First, we
consider the case, when one of the axions remains massless after moduli stabilization.
As explained in section 3, this still might be consistent with stabilizing the 
remaining moduli in a dS vacuum. In this case, we show that it is possible 
to find stable global strings in classification 
of~\cite{Polchinski}.\footnote{The remaining axion can receive a potential 
in the low-energy field theory from, for example, QCD instantons. 
Moreover, the non-existence of continuous global symmetries in string theory
implies that, eventually, each axion has to receive a potential. 
We will not 
discuss it in this paper.}
In fact, an open membrane studied in~\cite{Cosmic}
is an example of global strings. However, depending on exactly which axion 
remains massless, it might happen that open membranes become unstable 
under domain wall formation, whereas five-branes wrapped on a four-cycle
become stable global strings. Then we consider the case when all axions are stabilized.
An open membrane can no longer be stable because 
the axion which it is not periodic around it becomes massive. 
The problem of domain wall formations is partially resolved if the axion is charged
under an anomalous $U(1)$ and gauged away~\cite{CMP}. However, in the $E_8 \times E_8$ theory,
only one linear combination of axions can be gauged~\cite{Polchinski}.
This means that a string can be stable only if it is charged under this 
gauged linear combination and uncharged under all the remaining massive axions.
These strings can be formed as bound states of open membranes, five-branes
and solitonic strings.
It is natural to propose that in a generic heterotic M-theory 
compactification, it is always possible to create a bound 
state with the property stated above.  
Under some circumstances these bound states are stable Aharonov-Bohm strings.
However, we point out that their production might not be 
consistent with moduli stabilization. 
Therefore, one can expect
production of strings which are not seen by any scalar field in the spectrum. 
By Polchinski's conjecture they have be unstable and break on monopoles.
We estimate the monopole masses in heterotic M-theory and argue that 
these strings can have a sufficiently long lifetime.

There are important issues which we 
will not consider in this paper.
Fisrt of all, cosmic strings can be detected only if they are produced after inflation. 
We will not discuss it in this paper. 
Some reasons why various heterotic cosmic strings can be produced after 
inflation can be found in~\cite{Cosmic, Beckercosmic}. 
Furthermore, there can be additional instability sources 
besides domain wall formation and breakage on monopoles.
For example, strings can dissolve into flux or break on the visible or hidden 
brane. Unlike the instabilities studied in this paper, these other processes
cannot be studied in a universal manner and very much model dependent. 
%For this reason, we will not consider quasi-Aharonov-Bohm strings~\cite{Polchinski} 
%in this paper
%though they might be an important example of heterotic cosmic strings.

%%%%%%%%%%%%%%%%%%%%%%%%%%%%%%%%%%%%%%%%%%%%%%%%%%%%%%%%%%%%%%%%%%%%%%%%%%%%%%%%%%%%%%%%%%%%%%%%%%%%%%%

\section{Wrapped Five-Branes as Cosmic Strings and An\-iso\-tro\-pic Calabi-Yau Threefolds}

%%%%%%%%%%%%%%%%%%%%%%%%%%%%%%%%%%%%%%%%%%%%%%%%%%%%%%%%%%%%%%%%%%%%%%%%%%%%%%%%%%%%%%%%%%%%%%%%%%%%%%%%%%

\subsection{The Tension}

%%%%%%%%%%%%%%%%%%%%%%%%%%%%%%%%%%%%%%%%%%%%%%%%%%%%%%%%%%%%%%%%%%%%%%%%%%%%%%%%%%%%%%%%%%%%%%%%%%%%%%%%%%%%

Five-branes wrapped on four-cycles in Calabi-Yau threefold can 
either themselves be viewed as strings in four dimensions
or be ingredients of string bound states. To make 
sure that the strings of interest will have a relatively small 
tension we have to understand under what condition the four-cycle
on the which the five-brane is wrapped can be made small comparing to the 
Calabi-Yau scale.

We consider a five-brane wrapped on a four-cycle and parallel 
to the orbifold fixed planes.\footnote{Throughout the paper we will refer to one of these planes
as to the visible sector and to the other one as to the hidden sector.}
From the four-dimensional viewpoint such a configuration will look like a string.
The tension of this strings was evaluated in details in~\cite{Beckercosmic}
so we will be very sketchy. The tension behaves as
\begin{equation}
\mu \sim M_{11}^6 v_4,
\label{2.1}
\end{equation}
where $M_{11}$ is the eleven-dimensional Planck scale and $v_4$ is the 
volume of the four-cycle.
It is easy to realize that in order for this tension to satisfy the 
observational bound 
\begin{equation}
G\mu \sim 10^{-7}
\label{2.2}
\end{equation}
the size of the four cycle should be sufficiently smaller that the Calabi-Yau scale.
See~\cite{Beckercosmic} for details. It is possible to create such an object 
only if it is possible to stabilize some of the $h^{1,1}$ moduli in such a way that 
one or more cycles have a small area. This will be discussed
in the subsection 2.3. 

%%%%%%%%%%%%%%%%%%%%%%%%%%%%%%%%%%%%%%%%%%%%%%%%%%%%%%%%%%%%%%%%%%%%%%%%%%%%%%%%%%%%%%%%%%%%%%%%%%%%%%%%%%

\subsection{A Topological Obstruction}

%%%%%%%%%%%%%%%%%%%%%%%%%%%%%%%%%%%%%%%%%%%%%%%%%%%%%%%%%%%%%%%%%%%%%%%%%%%%%%%%%%%%%%%%%%%%%%%%%%%%%%%%%%%

There is a topological obstruction to having a five-brane wrapped
on a four-cycle in the bulk. It comes from the fact the Bianchi identity 
requires the existence of the flux $G_{(2,2,0)}$~\cite{Wittenstrong}.
The notation is that $G_{(2,2,0)}$ has two holomorphic, two antiholomorphic 
indices along Calabi-Yau and no indices along the interval. This flux provides 
a warping of the metric along the interval~\cite{Wittenstrong, CK2, CK3, Beckercosmic}
\begin{equation}
ds^2 =e^{-f(x^{11})}g_{\mu \nu}dx^{\mu} dx^{\nu}
+e^{f(x^{11})}( g_{CYmn}dy^m dy^n +dx^{11}dx^{11}),
\label{n2.3}
\end{equation}
where 
\begin{equation}
e^{f(x^{11})} = (1-x^{11}G)^{2/3}
\label{n2.4}
\end{equation}
and $G$ is given by
\begin{equation}
G= \int \omega \wedge G_{(2,2,0)}.
\label{n2.5}
\end{equation}
In general, a heterotic M-theory compactification contains five-branes wrapped 
on holomorphic two-cycles. To simplify language, in this subsection, we
will refer to five-branes wrapped on two-cycles as to three-branes and
to the five-brane wrapped on a four-cycle as to one-brane. In the absence of three-branes
in the bulk, $G$ in eq.~\eqref{n2.5} is given by~\cite{Wittenstrong}
\begin{equation}
G =\frac{-1}{8 \pi V}(\frac{\kappa_{11}}{4\pi})^{2/3}
\int \omega \wedge (tr {\cal F}^{(1)}\wedge {\cal F}^{(1)} -\frac{1}{2}tr {\cal R}\wedge {\cal R}),
\label{n2.51}
\end{equation}
where ${\cal F}^{(1)}$ is the instanton on the visible sector and $\omega$ is the Kahler form.
In the presence of the three-branes, 
the right hand side of eq.~\eqref{n2.51} will be modified by the cohomology class of the three-branes. 
Since the integral in~\eqref{n2.51} is an integer (up to normalization) and the integral of 
$\omega$ over any two-cycle is positive, it follows that if the flux $G_{(2,2,0)}$ is topologically 
non-trivial, the integral of $G_{(2,2,0)}$ over any four-cycle is non-zero. This means that there is 
a problem with wrapping a five-brane on a four-cycle. Indeed, on a worldvolume of the five-brane
there is a coupling 
\begin{equation}
\int B_{2} \wedge G_{(2,2,0)}.
\label{n2.50}
\end{equation}
This coupling can be understood if one compactifies one of the non-compact directions 
on a circle. Then one obtains a type IIA $D4$-brane on a four-cycle with the coupling 
\begin{equation}
\int A_{1} \wedge G_{(2,2,0)},
\label{n2.6}
\end{equation}
which is known to be there. From eq.~\eqref{n2.50} it follows that $G_{(2,2,0)}$
acts like a source for $B_2$ which has to be canceled. Otherwise, such a configuration 
is not allowed. 

In this paper, we will restrict ourselves to the case when $G_{(2,2,0)}$
is cohomologically trivial in some region in the bulk. Then the coupling eq.~\eqref{n2.50}
goes away and there is no obstruction to wrapping a five-brane on a four-cycle. Let us discuss
under what circumstances it can be achieved. The cohomology class of the $(2,2,0)$ component
of the flux is given by~\cite{HW2, Wittenstrong}
\begin{equation}
G_{(2,2,0)}(x^{11}) = (c_2 (V_1) -\frac{1}{2} c_2 (X)) \epsilon(x^{11}) 
+(c_2 (V_2) -\frac{1}{2} c_2 (X)) \epsilon(x^{11}-\pi \rho) 
+ \sum_I [W_I] \epsilon (x^{11}-x_I).
\label{n2.7}
\end{equation}
Here $c_2(V_1)$ and $c_2(V_2)$ are the second Chern classes of the vector bundles 
on the visible and hidden sectors respectively, $c_2(X)$ is the second Chern 
class of the Calabi-Yau threefold $X$,
$x_I$ is the position of the 
$I$-th three-brane whose cohomology class is $[W_I]$. The function $\epsilon(x)$ is defined
as $+1$ for $x > 0$ and $-1$ for $x < 0$. Let us pick a point along the interval where 
we would like to place a one-brane. Let $[W_1]$ be the total three-brane class to the left 
of this point and 
$[W_2]$ be the total three-brane class to the right. We have 
\begin{equation}
G_{(2,2,0)}=c_2(V_1) -c_2(V_2) +[W_1]-[W_2].
\label{n2.8}
\end{equation}
We want this to be zero. Therefore, we set
\begin{equation}
c_2(V_1) -c_2(V_2) +[W_1]-[W_2] =0
\label{n2.9}
\end{equation}
This equation should be supplemented by the usual anomaly cancellation condition~\cite{Wittenstrong}
\begin{equation}
c_2(V_1) + c_2(V_2) +[W_1] +[W_2] =c_2(X).
\label{n2.10}
\end{equation}
These two equations can be rewritten as
\begin{equation}
2c_2(V_1) + 2[W_1]  =c_2(X)
\label{n2.11}
\end{equation}
and 
\begin{equation}
2c_2(V_2) + 2[W_2]=c_2(X).
\label{n2.12}
\end{equation}
Thus, we are interested in compactifications where eqs.~\eqref{n2.11} 
and~\eqref{n2.12} are satisfied for some three-brane classes $[W_1]$ and $[W_2]$. 
The necessary condition for eqs.~\eqref{n2.11} and~\eqref{n2.12} to have a solution
is that $c_2(X)$ must be represented by an even four-form. If this condition is satisfied, 
then, in general, it should not be difficult to satisfy eqs.~\eqref{n2.11} and~\eqref{n2.12}
provided one can allow many three-branes in the bulk. 
It is important to note that allowing many three-branes in the bulk is consistent 
with stabilization of the interval in a phenomenological range~\cite{Cosmic}.
This will be reviewed in the next subsection. Let us now discuss the condition that
$c_2(X)$ must be even. We will simply present some examples of Calabi-Yau threefolds
with even $c_2(X)$ as an argument for possibility of existence of solutions of
eqs.~\eqref{n2.11} and~\eqref{n2.12} for some compactifications. 
Let us consider Calabi-Yau threefolds elliptically fibered over the Hirzebruch 
or del Pezzo surfaces. These manifolds have been used for analyzing 
heterotic GUT vacua in~\cite{Holomorphic, Instanton}. The second Chern class
$c_2(X)$ for such manifolds was computed by Friedman, Morgan and Witten in~\cite{FMW} 
and was found to be 
\begin{equation}
c_2(X) =\pi^* c_2(B) +11 \pi^* c_1(B)^2 +12 \sigma \cdot \pi^* c_1(B).
\label{n2.13}
\end{equation}
Here $c_1(B)$ and $c_2(B)$ are the first and the second Chern classes of the base $B$, 
$\pi$ is the projection map of $X$ to the base and $\sigma$ is the global section. 
The necessary condition of $c_2(X)$ to be even is that 
\begin{equation}
c_2(B) +11 c_1(B)^2
\label{n2.131}
\end{equation}
is an even number. For the Hirzebruch surfaces one has 
(see, for example, appendix B of~\cite{Holomorphic} and references therein 
for properties of the Hirzebruch and del Pezzo surfaces)
\begin{equation}
c_2(B)=4, \quad c_1(B)^2 =8. 
\label{n2.14}
\end{equation}
So $c_2(X)$ is an even class. For del Pezzo surfaces $dP_r$ one has
\begin{equation}
c_2(B) =3+r, \quad c_1(B)^2= 9+r.
\label{n2.15}
\end{equation}
Then it follows from eqs.~\eqref{n2.13} and~\eqref{n2.131} that $c_2(X)$ is even for any $r$. 
One can check that $c_2(X)$ is also even for more complicated Calabi-Yau threefolds 
with a non-trivial homotopy group which were used 
in~\cite{Volker1, Volker2, Donagi, Volker3} for obtaining heterotic Standard Model vacua. 

This shows that it is conceivable to obtain quasi-realistic compactifications
with $[G_{(2,2,0)}]=0$ in some region in the bulk. 
If we put a one-brane in this region, we obtain a consistent configuration.
Note that the warp-factor is unity in such a region and a one-brane is a BPS object. 
Also note that the flux $G_{(2,2,0)}=0$ is discontinuous across three-branes. 
Therefore, the warp-factor can be unity only in some region between two three-branes. 
At a generic position along the interval the warp-factor will still be non-trivial. 

Let us now discuss the case when a one-brane is placed in a region with non-trivial 
$G_{(2,2,0)}$. For consistency, the source coming from~\eqref{n2.50} has to cancel. 
The only possibility to do it is to allow membranes to end on a one-brane. 
These membranes will look like a string in four non-compact dimensions
and stretch along the interval. Their one end will be located on the one-brane 
and their other end will be located on either a three-brane or one of the orbifold planes. 
The number of membranes should be such that the source coming from the flux cancels. 
This membrane/five-brane configuration is totally consistent with string dualities. 
Indeed, upon compactifying one of the non-compact directions of this string bound state 
on a circle one obtains a type IIA $D4$-brane on a four-cycle with fundamental strings attached. 
This mechanism of canceling the source originating from the flux 
is totally analogous to the case of branes wrapping $S^5$ in AdS/CFT correspondence~\cite{Baryons}.
Due to the flux through $S^5$ in order to wrap a brane on $S^5$ one has to attach strings to the brane. 

From the four-dimensional viewpoint, the membrane/five-brane configuration looks 
like a string bound state. It is not BPS since its tension receives an extra contribution 
from the membranes. In there is only one membrane needed to cancel the flux, the tension is
\begin{equation}
\mu =\mu_5 + \mu_2 =\int_{4-cycle} d^4y \sqrt{-g} +\int dx^{11} \sqrt{-g},
\label{n2.19}
\end{equation} 
where the metric is given by eq.~\eqref{n2.3} and the integral over $dx^{11}$ 
is over the length of the membrane. The dependence of $\mu$ on $x^{11}$, up to a constant, 
is given by 
\begin{equation}
\mu \sim (1-x^{11}G)^{2/3}. 
\label{n2.20}
\end{equation}
This function does not have a minimum so this bound state has to move until 
it coincides with one of the three-branes of one of the orbifold planes. 
This is not surprising since this state is not BPS. Once a one-brane is on top 
of a three-brane or an orbifold plane the description in terms of the five-brane
on a four-cycle breaks down. Unfortunately, it is not known what happens 
when two M-theory five-branes come on top of each other. So if a one-brane 
coincides with a three-brane it is hardly possible to say what kind of object arises.
Things are slightly better when a one-brane coincides with an orbifold plane. 
The arising state should be interpreted as an instanton on $R^{1,3} \times X$. 
The second Chern class of this instanton should have two indices along $R^{1,3}$ and two indices
along $X$. It would be interesting to find an appropriate solution by, for example, 
assuming that the instanton is small and can be approximated by an instanton on $R^4$. 
However, it is very likely that if the scale of this instanton can be stabilized
it is due to the existence of compact directions which makes the analysis much more 
complicated. We will not consider this case in this paper leaving it for future research.

Let us summarize the results of this subsection. We argued that under some circumstances
it is conceivable to find a region where the flux is cohomologically trivial. Then there is
no obstruction to placing a five-brane wrapped on a four-cycle in this region. 
In this paper, we will assume that such a region can be found. 
On the other hand, if one places such a five-brane in a region with non-trivial flux, 
it tends to move until it coincides with a three-brane or an orbifold plane. 
In both case the original description in terms of a brane on four-cycle breaks down.
We will not pursue this issue in this paper. 

%%%%%%%%%%%%%%%%%%%%%%%%%%%%%%%%%%%%%%%%%%%%%%%%%%%%%%%%%%%%%%%%%%%%%%%%%%%%%%%%%%%%%%%%%%%%%%%%%%%%%%%%%%

\subsection{Stabilization of the $h^{1,1}$ moduli}

%%%%%%%%%%%%%%%%%%%%%%%%%%%%%%%%%%%%%%%%%%%%%%%%%%%%%%%%%%%%%%%%%%%%%%%%%%%%%%%%%%%%%%%%%%%%%%%%%%%%

In this subsection, we will discuss the possibility of stabilizing 
one or more two-cycles in a Calabi-Yau threefolds at a small area.
This will create an anisotropy which might be relevant for 
production of cosmic strings with a small tension 
by wrapping 
five-branes on four-cycles of a small volume. 
For simplicity, we will consider the case when $h^{1,1}=2$
though the generalization for any $h^{1,1}>1$ is straightforward.
Our aim is to create a potential energy that will tend to produce
an anisotropy of the Calabi-Yau threefold.
We consider the following system of moduli
\begin{equation}
Z_{\alpha}, T_{1}, T_2, Y_I, \quad I=1, \dots N, 
\label{2.7}
\end{equation}
where $Z_{\alpha}$ are the complex structure moduli, whose actual number will be irrelevant,
$T_1$ and $T_2$ are the two Kahler structure moduli, $Y_I$ are the moduli 
of five-branes wrapped on the same isolated genus zero holomorphic curve. For simplicity,
we will assume that we can choose the basis of curves in such a way that the volume
of the cycles in different homology classes are controlled exactly by $ReT_i$. 
All the five-brane will wrap the curve whose volume is controlled by $ReT_1$.
The precise structure of the $T_i$ moduli is as follows~\cite{LOW4, LOSW5}
\begin{equation}
T_i=Rb_i +ip_i.
\label{2.8}
\end{equation}
Here $R$ is the size of the interval, $b_i$ are the Kahler moduli of the Calabi-Yau 
threefold and $p_i$ are internal components of the three-form potential.
The moduli $b_i$ are not independent, 
\begin{equation}
\sum_{i,j,k}d_{ijk}b_ib_jb_k =6,
\label{2.9}
\end{equation}
where $d_{ijk}$ are the triple intersection numbers.
The five-brane moduli are defined as follows~\cite{Derendinger, Moore} 
\begin{equation}
Y_I=\frac{y_I}{\pi \rho}ReT_1 +i(a_I +\frac{y_I}{\pi\rho}ImT_1).
\label{2.10}
\end{equation}
Here $y_I$ is the actual positions of the $I$'th five-brane, $\pi\rho$ is the
reference length of the interval
and $a_I$ is the axions on the worldvolume of the $I$'th five-brane.
This system of moduli should be supplemented by the volume multiplet
\begin{equation}
S=(V+\dots)+i \sigma,
\label{2.11}
\end{equation}
where $V$ is the Calabi-Yau volume in the middle of the interval, $\sigma$ is 
the axion dual to the antisymmetric tensor and by the ellipsis we indicate 
that the real part of $S$ gets modified by in presence of 
five-branes~\cite{Derendinger, Moore}. We will assume that this modification
is small enough and can be ignored.
All moduli are dimensionless and normalized to the the reference scales
$v_{CY}^{1/6}\sim (10^{16} GeV)^{-1}$ and $\pi \rho \sim (10^{15}GeV)^{-1}$.
By construction, $ReY_I<ReT_1$. To obtain the four-dimensional coupling constants
in a phenomenological range, $V$ and $R$ have to be stabilized at
(or be slowly rolling near) a value of order one.
As will be discussed later, stability of various types of cosmic strings
depends whether or not there is a potential for the $S$ modulus.
Thus, we will postpone our discussion of stabilization of $S$
till section 4. 
Let us also point out that in a generic compactification
the system~\eqref{2.7} should also be supplemented by the vector bundle 
moduli. For simplicity, we will ignore them.

We want to understand how to stabilize the system~\eqref{2.7}
in an anti de Sitter (AdS) vacuum with $b_2 << b_1$. Stabilization of $S$ and de Sitter 
vacua will be discussed in the next section. 
The Kahler potential of this system is given by the standard expression
\begin{eqnarray}
\frac{K}{M^2_{Pl}} & = &-\ln\left(-i\int \Omega(Z) \wedge \bar\Omega(\bar Z)\right)
-\ln \left(\frac{1}{6}\sum_{ijk}d_{ijk}(T_i+\bar T_i)(T_j+\bar T_j)(T_k+\bar T_k)\right) \nonumber\\
& &+\sum_I K_1 (Y_I+\bar Y_I)^2,
\label{2.12}
\end{eqnarray}
where 
\begin{equation}
K_1=\frac{\tau_5}{(T_1+\bar T_1)(S+\bar S)}
\label{2.13}
\end{equation}
and $\tau_5$ is the five-brane tension. 
It was proposed in~\cite{Derendinger, Moore}
that the $Y$-moduli and the $S$ moduli Kahler potential forms on logarithm
\begin{equation}
-\ln \left(S+\bar S -\sum_I \frac{\tau_5(Y_I+\bar Y_I)^2}{T_1+\bar T_1} \right).
\label{2.14}
\end{equation}
However, in this paper, we will assume that the $Y$-dependent corrections are small enough 
and it is accurate to expand the logarithm keeping the quadratic terms in $Y$.
The superpotential for the system will be 
\begin{equation}
W=W_f+W_{np}.
\label{2.15}
\end{equation}
We will assume that the $(3,0,1)$ $G$-flux is turned on. It produces the superpotential 
for $Z_{\alpha}$ of the form~\cite{GVW}
\begin{equation}
W_f =\frac{M^2_{Pl}}{v_{CY}\pi \rho} \int d x^{11}\int_{CY}G \wedge \Omega.
\label{2.16}
\end{equation}
The coefficient in front is moduli independent as it is just $M_{11}^9$.
This superpotential is expected, generically, to stabilize all the
complex structure moduli. We will assume that it is the case.
We will also assume that the complex structure moduli are stabilized at slightly
higher scale that all the other moduli so that $W_f$ can be considered constant.
The non-perturbative superpotential $W_{np}$~\cite{Wittensuper, BBS, Moore, Lima1, Lima2, BDO1, BDO2}
is, approximately,
\begin{equation}
W_{np}= W_{np}(T_1, Y_I) +W_{np}(T_2).
\label{2.17}
\end{equation}
In the presence of the five-branes, the leading contribution
to the $T^1$ superpotential is due to open membranes stretched between 
the adjacent branes
\begin{equation}
 W_{np}(T_1, Y_I) \sim A_1 e^{-\tau_1 Y_1} + A_2 e^{-\tau_1(Y_2-Y_1)}
+ \dots A_{N+1} e^{-\tau_1 (T_1-Y_N)}.
\label{2.18}
\end{equation}
The coefficients $A_1, \dots A_{N+1}$ depend, in general, on the complex structure
and vector bundle moduli. For the purposes of this paper they can be taken to be constants. 
The coefficient $\tau_1$ was estimated in~\cite{BO} and found to be of order $250$.
Since we do not have any five-branes wrapped on cycles whose area is controlled 
by $ReT_2$, the term $W_{np}(T_2)$ is given by
\begin{equation}
W_{np}(T_2) \sim e^{-\tau_2 T_2}.
\label{2.19}
\end{equation}
First, let us consider
\begin{equation}
D_{T_1}W=0, \quad D_{Y_I}W=0
\label{2.20}
\end{equation}
These equations were analyzed in some detail in
~\cite{Cosmic}.\footnote{For brevity, we will not
discuss the imaginary parts of the moduli. Their stabilization 
was discussed in~\cite{BO, Raise}.}
The properties of the solution are the following.
\begin{enumerate}
\item 
The distances between the adjacent branes are approximately the same.
\item
Due to existence of the constant (for our purposes) superpotential $W_f$,
this distance will be proportional to $-\ln(|W_f|/\tau_1)$.
\end{enumerate}
This is easy to understand intuitively.
In the absence of $W_f$, the superpotential~\eqref{2.15}, \eqref{2.18}
will stabilize the relative distance between the adjacent branes. For example, 
if there is only one five-brane in the bulk it will be stabilized approximately 
in the middle of the interval. However, the overall interval modulus will run away.
The presence of $W_f$ will stabilize the remaining run-away modulus. If the number 
of five-branes is $N$, the relative distance is approximately $\frac{Rb_1}{N+1}$.
Eqs.~\eqref{2.20} lead to the following equation for the relative distance
\begin{equation}
e^{-\tau_1 \frac{Rb_1}{N+1}} \sim \frac{|W_f|}{\tau_1}.
\label{2.21}
\end{equation}
In order this equation to have a solution, the right hand side
has to be less than one. First, $W_f$ is quantized in units of
$(\frac{\kappa_{11}}{4\pi})^{2/3}$. Therefore, in the limit of
large volume and large interval, $|W_f|$
is less than one. Second, the order of magnitude of $W_f$ might be
reduced by Chern-Simons invariants~\cite{Gukovhet}. Third, $\tau_1$
is much greater than one. This guarantees that, generically, the
right hand side in eq.~\eqref{2.21} is less than one.
If $N$ is big enough (a simple estimate shows that $N \sim 20$
should suffice) one can always obtain
\begin{equation}
Rb_1 \sim 1.
\label{2.22}
\end{equation}

Now let us consider the equation
\begin{equation}
D_{T_2}W=0.
\label{2.23}
\end{equation}
We have 
\begin{equation}
e^{-\tau_2 Rb_2} \sim \frac{|W_f|}{\tau_2}.
\label{2.24}
\end{equation}
For $\tau \sim 10^{-3}-10^{-2}$ and generic $|W_f|$ or order, for example,
$10^{-2}$ in eleven-dimensional Planck units~\cite{BO} (we are assuming 
that the scale of $W_{np}$ is set by the eleven dimensional Planck scale)
this equation can have a solution only if
\begin{equation}
Rb_2 <<1.
\label{2.25}
\end{equation}
For generic triple intersection numbers $d_{ijk}$, 
our solution 
\begin{equation}
Rb_1 \sim 1, \quad Rb_2 <<1
\label{2.26}
\end{equation}
is consistent with having 
\begin{equation}
R \sim 1, \quad b_1 \sim 1, \quad b_2 <<1.
\label{2.27}
\end{equation}
This analysis provides a mechanism for creating an anisotropy of
the Calabi-Yau threefold. Creating just one cycle with a small area
might not be sufficient to create a four-cycle whose
volume is small enough to satisfy the bound estimated in~\cite{Beckercosmic}. 
However, it is straightforward to generalize this analysis
for the case $h^{1,1}>2$. By varying the number of five-branes 
wrapping cycles in various homology classes one can create 
enough anisotropy for existence of four cycles with a sufficiently small volume. 
A five-branes wrapped on such a cycle will look like a string with a small 
tension in Minkowski space. 

%%%%%%%%%%%%%%%%%%%%%%%%%%%%%%%%%%%%%%%%%%%%%%%%%%%%%%%%%%%%%%%%%%%%%%%%%%%%%%%%%%%%%%%%%%%%%%%%%%%%%%%%

\section{Volume Stabilization and de Sitter Vacua}

%%%%%%%%%%%%%%%%%%%%%%%%%%%%%%%%%%%%%%%%%%%%%%%%%%%%%%%%%%%%%%%%%%%%%%%%%%%%%%%%%%%%%%%%%%%%%%%%%%%%%%%

As explained in the introduction,
in order to understand what kind of strings can be produced and whether or not they are stable, 
it is important to formulate the set-up for moduli stabilization.
In this section, we will complete this set-up by discussing various mechanisms
for stabilizing the volume multiplet. 

The most common way to stabilize the volume multiplet is to allow the hidden sector gaugino
to condense. This produces the superpotential of the 
form~\cite{Horava, Ovrut1}
\begin{equation}
W_g = he^{-\epsilon S +\dots}.
\label{3.1}
\end{equation}
Here $h$ is the coefficient whose scale is set by $M_{11}^3$, $\epsilon$ is given by~\cite{Ovrut1}
\begin{equation}
\epsilon =\frac{2\pi}{b_0 \alpha_{GUT}}, 
\label{3.2}
\end{equation}
where $b_0$ is the coefficient of the one-loop beta function and the ellipsis
indicate the threshold corrections depending on other moduli due to fact that 
the Calabi-Yau volume is warped along the interval. 
They will not be important for us. 
Adding $W_g$ to $W$ in eq.~\eqref{2.15} stabilizes the volume~\cite{BO, Gukovhet}.
As the result, the system of moduli~\eqref{2.7} supplemented by the volume multiplet 
$S$ (and, in principle, by the vector bundle moduli, which, for simplicity, have been ignored)
can be stabilized in a supersymmetric AdS vacuum.

In order to raise this AdS vacuum to a metastable dS vacuum we will assume that the 
low-energy gauge group contains a $U(1)$ factor which is anomalous. 
This anomaly is canceled by the four-dimensional version of the Green-Schwarz
mechanism~\cite{Witten32, Dine}. The axion, in this case, transforms under $U(1)$ as
\begin{equation}
\sigma \to \sigma +\l.
\label{3.3}
\end{equation}
As was pointed out in~\cite{Blumen, Polchinski}, in $E_8 \times E_8$ compactifications,
the existence of the anomalous $U(1)$ implies that the 
imaginary parts of the $h^{1,1}$ moduli are also charged.
The reason is the following. Due to $E_8$ properties, the anomaly in the $U(1)$ 
generator $T$ is a linear combination of
\begin{equation}
\int (Tr T{\cal F})^3, \quad 
\int (TrT^2)(TrT{\cal F})(Tr {\cal F}^2), \quad 
\int (TrT^2)(TrT {\cal F}) (Tr {\cal R}^2).
\label{3.4}
\end{equation}
This means that the anomaly $T$ exists only if the integral 
of $Tr {\cal F}$ over at least one two-cycle is non-zero. 
However, under the same conditions,
the anomalous $U(1)$ gauge field $A$ couples to at least one imaginary part
of the $h^{1,1}$ moduli. This coupling will come
from the term (after we perform the dimensional reduction on $S^1/Z_2$~\cite{LOW4})  
\begin{equation}
\int H\wedge *H, 
\label{3.5}
\end{equation}
where 
\begin{equation}
H=dB_2- (\omega_{YM}-\frac{1}{2}\omega_L).
\label{3.6}
\end{equation}
The simplest way to proceed is to dualize $dB$ by introducing a six-form $B_6$ 
\begin{equation}
dB_6 =*dB.
\label{3.7}
\end{equation}
Then eq.~\eqref{3.5} will contain a term
\begin{equation}
\int B_6 \wedge (tr {\cal F}^2 -\frac{1}{2}{\cal R}^2).
\label{3.8}
\end{equation}
Upon the dimensional reduction to four dimensions this will produce a term
\begin{equation}
\int p_2 \wedge dA  = \int d^4x p\partial_{\mu}A^{\mu},
\label{3.9}
\end{equation}
provided the integral of ${\cal F}$ over some two-cycle is not zero.
Here $p_2$ is the antisymmetric tensor dual to the imaginary part $p$ of some 
$h^{1,1}$ modulus. Coupling~\eqref{3.9} guarantees that the axion $p$ is 
also charge under the anomalous $U(1)$~\cite{Dine}. If $h^{1,1}$ is bigger than one, 
in general, more than one axions $p_i$ will be charged under $U(1)$.

The anomalous $U(1)$ also leads to FI terms~\cite{Dine}.
In heterotic M-theory their scale is approximately the same as 
that of $\frac{W_f^2}{M_{Pl}^2}$~\cite{Raise}
so that they can be used to raise a supersymmetric AdS vacuum to a dS vacuum~\cite{BKQ}.
Remarkably, the fact that there is always more than one axion charged 
under $U(1)$ implies that even in the absence of a superpotential for the
multiplet $S$, the volume $V$ can be stabilized in a dS vacuum if FI terms are present. 
For simplicity, let us consider the case $h^{1,1}=1$. We will not be very careful
about all the factors. They were estimated in~\cite{Raise}.
The transformation law~\eqref{3.3} implies that the Kahler potential for the $S$ multiplet has to 
be of the form
\begin{equation}
K(S)=-\ln(S+\bar S+ {\cal V}),
\label{3.10}
\end{equation}
where ${\cal V}$ is the anomalous $U(1)$ vector superfield. 
The term
\begin{equation}
\int d^4 x d^4 \theta K(S)
\label{3.11}
\end{equation}
produces, among other terms, the FI term 
\begin{equation}
\int d^4 x \frac{D}{V},
\label{3.12}
\end{equation}
where $D$ is the auxiliary field. Then the contribution to the potential 
energy is 
\begin{equation}
\frac{g^2}{V^2}.
\label{3.13}
\end{equation}
The gauge coupling constant $g^2$ is also moduli dependent
\begin{equation}
g^2 \sim \frac{1}{V+\gamma},
\label{3.131}
\end{equation}
where $\gamma$ stands for the threshold corrections whose precise form 
is not important. They can be found, for example, in~\cite{LOW4}. 
As was explained, the axion $p$ also transforms the same way which means that the 
Kahler potential for the $T$-modulus is of the form
\begin{equation}
K(T)=-\ln(T+\bar T+ {\cal V}).
\label{3.14}
\end{equation}
This implies that there is a correction
to the FI potential energy of the form
\begin{equation}
\frac{g^2}{R^2}.
\label{3.15}
\end{equation}
If $h^{1,1}$ is grater than one the denominator $R^2$ is replaced with 
some more complicated function of $ReT_i$.
These two contributions imply that the volume-dependent
FI contribution to the potential energy is of the form
\begin{equation}
U_{FI} = \frac{B}{V^2(V+\gamma)}+\frac{A}{V+\gamma}.
\label{3.16}
\end{equation}
The coefficient $A$ depends on 
the $h^{1,1}$ moduli which were stabilized in the previous section and can be considered constants.
Thus, approximately, dynamics of $V$ is governed by the
potential\footnote{The FI terms will modify equations of motion
for the $h^{1,1}$ moduli and the five-branes. However, since the order of
magnitude of FI terms is the same as that of the fluxes, a
solution for these moduli will still exist. It is also possible to show that 
their Kahler covariant derivatives will be shifted by terms proportional by $\frac{1}{\tau_1}$
or $\frac{1}{\tau_2}$ which is much less than one.}
\begin{equation}
U(V) =e^{K(S)} [G^{-1}_{S \bar S} D_S W D_{\bar S} \bar W -3 W \bar W]+U_{FI}.
\label{3.17}
\end{equation}
The supergravity term in~\eqref{3.17} gives just $\frac{-2}{V}$ up to terms quartic
on the five-brane positions which will be assumed to be small enough and neglected. 
Thus, the potential becomes
\begin{equation}
U(V)=-\frac{2}{V}+\frac{B}{V^2(V+\gamma)}+\frac{A}{V+\gamma}.
\label{3.18}
\end{equation}
Under some conditions on parameters $A, B$ and $\gamma$ this function can have
a dS vacuum. See fig. 1.
This analysis shows that even in the absence of any superpotentials for the $S$ multiplet, 
one can stabilize all heterotic moduli in a dS vacuum. 
\begin{figure}
\epsfxsize=3.5in \epsffile{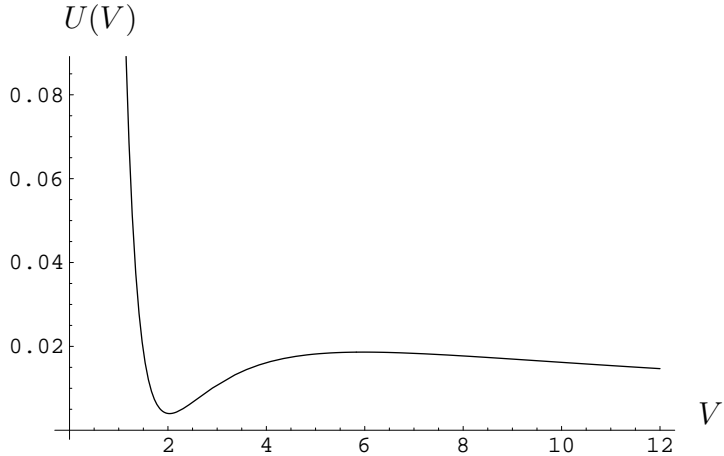}
\begin{picture}(30,30)
\put(-235,165){$U(V)$}
\put(5, 15){$V$}
\end{picture}
\caption{Potential $U(V)$ (appropriately normalized) for $A=2.26$,
$B=1$, $\gamma=0.5$. There is a dS minimum.}
\end{figure}

There is one important comment we have to make before finishing this section.
In the presence of the anomalous $U(1)$ the non-perturbative 
superpotentials~\eqref{2.18}, \eqref{2.19}, \eqref{3.1} do not seem gauge 
invariant. To keep the gauge invariance they must be multiplied by some
power a matter field charged under $U(1)$. Let us denote this field by
$Q=re^{i\phi}$. Then the variation of the axion will be canceled
by the variation of the phase $\phi$. To make sure that the non-perturbative
superpotentials stabilize the $h^{1,1}$ moduli, $r$ should receive a vacuum 
expectation value. Thus, the system of moduli discussed in this and in the previous 
sections has to be supplemented by the field(s) $Q$. The radius of $Q$, $r$, 
receives a very complicated potential coming from the both the $F$- and $D$-terms.
We will not study it in this paper. Generically, this potential can stabilize 
$r$ at a non-zero value. We will assume that it is the case and $r$ is stabilized at 
a value of order one. 

 %%%%%%%%%%%%%%%%%%%%%%%%%%%%%%%%%%%%%%%%%%%%%%%%%%%%%%%%%%%%%%%%%%%%%%%%%%%%%%%%%%%%%%%%%%%%%%%%%%%%%%%%

\section{Axions and Cosmic Strings}

%%%%%%%%%%%%%%%%%%%%%%%%%%%%%%%%%%%%%%%%%%%%%%%%%%%%%%%%%%%%%%%%%%%%%%%%%%%%%%%%%%%%%%%%%%%%%%%%%%%

\subsection{Global Heterotic Strings}

%%%%%%%%%%%%%%%%%%%%%%%%%%%%%%%%%%%%%%%%%%%%%%%%%%%%%%%%%%%%%%%%%%%%%%%%%%%%%%%%%%%%%%%%%%%%%%%%%%

After we introduced various moduli stabilizing potentials we can proceed 
to discuss what kind of strings can be found under different circumstances.
There are three natural candidates for cosmic strings in heterotic M-theory:
open membranes, five-branes wrapped on four-cycles and solitonic strings.
The latter can exist by Kibble's argument~\cite{Kibble1} because at least one
field charged under the anomalous $U(1)$ receives a vacuum expectation 
value as pointed out in the previous section. All these three types of strings 
can potentially have a small tension. An open membrane can have a small 
tension due to the possibility of stabilizing a five-brane wrapped on an isolated genus
zero curve close to the visible sector. This was studied in~\cite{Cosmic}.
A five-brane wrapped on a four-cycle can have a small tension due to the possible 
Calabi-Yau anisotropy as explained in section 2. A solitonic string can have a small 
tension because the scale of both the potential and kinetic energy for the charged 
fields is much less than the four-dimensional Planck scale. The scale of the 
potential energy is of order $\frac{W_f^2}{M_{Pl}^2}$. 
This quantity is naturally less than $M_{Pl}^4$ 
by many orders of magnitude. The scale of the kinetic energy is much less that the Planck scale
because the matter fields in heterotic M-theory originate from the orbifold sector.
When we normalize the kinetic term the vacuum expectation value of the 
matter field of interest will be much less than $M_{Pl}$. These arguments suggest that 
the tension of a solitonic string must be much less than one in four-dimensional 
Planck units. 

The most dangerous source of instabilities of heterotic cosmic strings is
formation of axion domain walls of non-zero tension~\cite{Vilenkin}.
All three types of strings mentioned above are boundaries of
axion domain walls. Once the corresponding axion receives a mass
the domain wall will make the string rapidly collapse.
A solitonic string is bounded by a domain wall of the phase $\phi$ of the field $Q$.
This is a standard field theory result. To have a solitonic string solution
means that far away from the string the field $\phi$ is an angular variable.
Thus, as we go around the string 
\begin{equation}
\int d\phi =2\pi,
\label{4.1}
\end{equation}
meaning that there is a domain wall bounded by the string. 
An open membrane, like a fundamental string in the weakly coupled $E_8 \times E_8$
theory is bounded by a domain wall of the axion $\sigma$~\cite{Wittencosmic}.
For completeness, let us present this argument following 
Witten~\cite{Wittencosmic}.
Let us say that we have a string stretched in the $x^3$ direction
and localized at $x^1=x^2=0$.
Let $\Gamma$ be a contour around the string and $\Sigma$ be a surface bounded 
by $\Gamma$. Then we have
\begin{equation}
\int_{\Gamma}d \sigma =\int_{\Gamma} *H=
\frac{1}{2}\int_{\Sigma} dx^1 dx^2 \partial^{\mu}H_{\mu 30}.
\label{4.2}
\end{equation}
Since a fundamental string in the weakly coupled theory and an open membrane
in heterotic M-theory couple to $H$, the equations of motion for $H$ implies
\begin{equation}
\partial^{\mu}H_{\mu 30} \sim \delta (x^1) \delta (x^2).
\label{4.3}
\end{equation}
Due to the delta-functional source, the integral~\eqref{4.2} is non-zero
meaning that $\sigma$ is not periodic as one goes around the 
string.
The next object of interest is a five-brane wrapped on a four cycle.
To simplify our discussion in this section we will consider 
the case $h^{1,1}=1$. Though this choice is not quite consistent with 
section 2, this is not very important as
the generalization of what follows for an arbitrary $h^{1,1}$ is straightforward.
This five-brane is charged under the axion $p$ which is the imaginary part 
of the $h^{1,1}$ modulus. In the case $h^{1,1}>1$, the five-brane will 
couple to the linear combination of $h^{1,1}$ axions associated to the two-form Poincare dual 
to the four-cycle on which it is wrapped.
Let us show it. Since the five-brane is magnetically charged under $H$ in ten 
dimensions, we have~\footnote{We compactified the theory on the interval 
and consider the three-form strength $H$ instead of the four-form 
strength $G$.} 
\begin{equation}
\int_{\Sigma_3} H \neq 0, \quad \Sigma_3 =\Sigma_2 \times \Gamma, 
\label{4.4}
\end{equation}
where $\Gamma$ is the contour around the string and $\Sigma_2$ 
is the two-cycle dual to the one on which the five-brane is wrapped.
Upon the dimensional reduction
\begin{equation}
H=dp \wedge \omega, 
\label{4.5}
\end{equation}
where $\omega$ is the Kahler form. Since
\begin{equation}
\int_{\Sigma_2} \omega =1,
\label{4.6}
\end{equation}
we obtain
\begin{equation}
\int_{\Sigma_3} H =\int_{\Gamma}dp \neq 0.
\label{4.7}
\end{equation}
This shows that the five-brane couples to the axion $p$.

Thus, all three types of strings have a potential instability 
after we stabilize the moduli.
To prevent an axion domain wall formation, the axion has to be 
charged under an anomalous $U(1)$ gauge group~\cite{CMP}
as in eq.~\eqref{3.3}. Then the axion is, effectively,
gauged away from the spectrum. 
However, as pointed out in~\cite{Polchinski} and reviewed in
the previous section, in the $E_8 \times E_8$ theory 
there are always more than one axions charged under the anomalous 
$U(1)$. So one can gauge away only one linear combination.
All that indicates that it is important to understand what kind of axion 
potentials can arise in the process of moduli stabilization
which was discussed in the previous two sections.
Let us understand which stable strings can be found in the set-up of~\cite{Cosmic} 
when the Kahler structure moduli are stabilized by non-perturbative effects 
but the volume multiplet receives no superpotential.
For this we have to understand in detail which axions receive a potential.
In the simplest case $h^{1,1}=1$ there are three axions
$\sigma, p$ and $\phi$. 
In the presence of the anomalous $U(1)$, which is always
assumed throughout the paper, all three of them are charged under it.
This means that one linear combination of the axions can be gauged away. Without loss
of generality, we can assume that this linear combination is 
$\sigma+ p +\phi$. Therefore, no domain walls of the axion 
$\sigma+ p +\phi$ can be formed.
The non-perturbative superpotential for the $h^{1,1}$ modulus
gives a potential for the linear combination of $p$ and $\phi$. 
Without loss of generality, we can assume that this linear combination is
$p-\phi$. On the other hand, since, by construction, no non-perturbative
superpotential for the volume multiplet is turned on, the uncharged linear 
combination of $\sigma$ and $\phi$, which we will take to 
be $\sigma-\phi$
has no potential.
The outcome is that $\sigma+ p +\phi$ is eaten by 
the broken $U(1)$ vector field, $p-\phi$ is massive and $\sigma-\phi$ 
is massless. As the result, a solitonic string is unstable 
because the massive field $p-\phi$ is not periodic as we go around the string.
For the same reasons, a five-brane on a for-cycle is also unstable.
However, an open membrane is stable because the only field which is not gauged and has a jump 
around it is $\sigma-\phi$ which is massless. In other words, 
in the given model, an open membrane studied in detail in~\cite{Cosmic}
is stable because it is a global string in Polchinski's classification~\cite{Polchinski}.  
In this model, there is one more global string. 
One can have a bound state of one $p$-string and one $\phi$ string
so that $p-\phi$ will be periodic as we go around the bound state. 
The massless field $\sigma -\phi$ has a jump as we go around it
so this string is also a global string.
On the other hand, a bound state of one open membrane and one solitonic string 
is unstable since the massive field $p-\phi$ is not periodic around this bound state.
Similarly, a bound state of an open membrane and a five-brane is also unstable.

Let us emphasize that the analysis of stability of various strings presented above 
depended on the exact structure of the potential for various axions.
One can imagine the situation when the volume multiplet receives a superpotential
due to a gaugino condensate in the hidden sector but
one or more of the $h^{1,1}$ moduli receives no superpotential. 
In this case, the real parts of the corresponding $h^{1,1}$ moduli
might be possible to stabilize by FI-terms and certain linear combinations
of the $h^{1,1}$ axions and the phase $\phi$ will be massless.
Now open membranes will bound non-zero tension domain walls and will be unstable, 
however five-branes on four-cycles will now be stable global strings.
In any case, if after moduli stabilization one or more axions
remain massless, stable global strings can be produced. 

%%%%%%%%%%%%%%%%%%%%%%%%%%%%%%%%%%%%%%%%%%%%%%%%%%%%%%%%%%%%%%%%%%%%%%%%%%%%%%%%%%%%%%%%%%%%%%%%%%%%

\subsection{Aharonov-Bohm Heterotic Strings}

%%%%%%%%%%%%%%%%%%%%%%%%%%%%%%%%%%%%%%%%%%%%%%%%%%%%%%%%%%%%%%%%%%%%%%%%%%%%%%%%%%%%%%%%%%%%%%%%%%%

In this and the next subsections, we will consider the case when all moduli 
multiplets receive a superpotential. As the result, after moduli 
stabilization there are no massless pseudoscalars left. 
The strings which were global and, thus, stable in the previous section will 
now bound domain walls of non-zero tension and quickly collapse. 
In the example considered above, we have the axion $\sigma+p+\phi$ eaten 
by the massive $U(1)$, whereas the $p-\phi$ and 
$\sigma-\phi$\footnote{In this discussion, we are ignoring the axion 
dependent threshold corrections to the gaugino condensation superpotential. 
Taking them into account would make our analysis more complicated 
without introducing any conceptual novelty.}
are both massive.
An open membrane is now unstable since it bounds a domain wall 
of the field $\sigma-\phi$ which is now massive. Similarly, a generic bound
state of the three types of strings will also be unstable.
A candidate for a stable string should be uncharged under the both massive fields.
The only field which is allowed not to be periodic around such a candidate is
$\sigma+p+\phi$ which is gauged. In the given example, it is easy to propose
a bound state with this property. The bound state of one open membrane, 
one five-brane and one solitonic string, all properly oriented, 
can respect periodicity of both $\sigma-\phi$ and $p-\phi$. 
The only non-periodic field around this bound state is $\sigma+p+\phi$.
This string is stable under domain wall formation. 
It is natural to propose that in any heterotic M-theory compactification,
regardless of details, it is always possible to construct a bound
state which is charged only under the linear combination of axions which is gauged.
Therefore, all such strings are potentially stable.

However, there is one more universal instability for all heterotic cosmic strings,
namely, breakage on monopoles. 
In~\cite{Polchinski}, Polchinski conjectured that in string theory every 
potential decay should be allowed.
This implies that
in the absence of any massless particles the only stable strings are Aharanov-Bohm strings
which have an Aharonov-Bohm phase with respect to a 
particle (more generally, a collection of particles)
neutral under the low-energy 
gauge group. The conjecture implies that the bound
state constructed above 
should be unstable unless it is an Aharonov-Bohm string. Breaking on monopoles 
is a natural decay process. 
In general, a concrete decay mechanism can be 
very complicated. In our case, the mechanism is qualitatively simple.
To illustrate it, 
let us consider first a toy example of the Abelian Higgs model
(for a review see, for example, \cite{Kibble2}).
In this model, we have a $U(1)$ vector field $A$ coupled to a complex scalar $f$ with the potential 
of the form $(|f|^2-\eta^2)^2$. The equations of motion will support a string solution
with a finite tension. The $U(1)$ is broken in the presence of this solution
and the phase of $f$ is gauged. The string contains a magnetic flux tube with quantized flux
\begin{equation}
\int F =\int_{S^1_{\infty}} A = \frac{2\pi n}{e}.
\label{4.8}
\end{equation}
Let us now assume that the $U(1)$ comes 
after we break some non-abelian gauge group so that the theory also has monopoles.
In the presence of the string the $U(1)$ is broken. Therefore, there is no long-range 
force and the monopoles have to be confined. Since a solitonic string contains a flux tube
it can end on monopoles, thus confining them. This process prevents strings from growing 
to cosmic sizes. 
In heterotic Standard Model-like models one also gets monopoles after
the GUT gauge group is broken to the Standard Model by Wilson lines. 
So the situation is not very different from the Abelian Higgs Model.
Like a solitonic string, a $(\sigma, p, \phi)$-bound state supports a magnetic flux.
Therefore, it can break on monopoles. In principle, one can imagine that 
there are no monopoles in our theory with minimal charge. Then only two or more 
strings can end on a monopole. However, Polchinski conjectured in~\cite{Polchinski}
that in string theory the charge quantization is always saturated.

Polchinski's conjecture says that a $(\sigma, p, \phi)$-bound state
is stable only if it is an Aharonov-Bohm string. That is, if there exist
particles charged under the anomalous $U(1)$ which pick up a phase 
$\frac{2\pi}{q}, q>1$ around the string. Let us see if a $(\sigma, p, \phi)$-bound state
can be an Aharonov-Bohm string. 
%We will assume that in our theory there are particles
%charged only under the anomalous $U(1)$. 
Recall that we gauge a linear combination of axions.
In a generic compactification, it is very likely that this combination 
will have a charge $q$ greater than one. Any $(\sigma, p, \phi)$-bound state of interest
is charged only under this linear combination. 
Then, if there is a particle of charge one under the anomalous $U(1)$, 
it will pick up a phase $\frac{2\pi}{q}$ as desired.  
Indeed, an axion $a$ of charge $q$ couples to the $U(1)$ through the term in the action
$(\partial_{\mu}a-qA_{\mu})^2$~\cite{Dine}. 
To minimize the energy of the string we should set
\begin{equation}
\partial_{\mu} a-qA_{\mu}=0.
\label{4.9}
\end{equation}
As we go around the string, the axion $a$ changes by one. This means that far away from 
the core of the string, $a$ is the angular variable which we denote by $\theta$. 
To satisfy~\eqref{4.9}, we have
\begin{equation}
A_{\theta} \sim \frac{1}{q}.
\label{4.10}
\end{equation}
This means that any particle of charge one will pick up a phase $\frac{2\pi}{q}$ as
in the usual Aharonov-Bohm effect.
Because of this argument, one can think it is very likely that a $(\sigma, p, \phi)$-bound state
can be Aharonov-Bohm and, thus, stable. Unfortunately, there is a serious obstruction 
against it coming from the process of moduli stabilization. As was discussed, 
to turn on various non-perturbative superpotentials, at least one charged under the 
anomalous $U(1)$ scalar field should receive a vacuum expectation value. 
This field has to be single-valued around the string which is possible only if
its charge divides $q$. This is a strong restriction which is unlikely to be satisfied 
in a generic compactification. However,  
if it is satisfied and there is another 
particle (or collection of particles) charged only under the anomalous $U(1)$
whose charge is not divisible by $q$, then a $(\sigma, p, \phi)$-bound state 
is a stable Aharonov-Bohm string. Otherwise, $q$ $(\sigma, p, \phi)$-strings should form 
a bound state to make sure that a particle with a non-zero vacuum expectation value is single-valued
around the string.
This bound state, in principle, can still be an Aharonov-Bohm string if there is
a particle with a fractional charge
which does not receive a vacuum expectation value.
%This bound state can be a quasi-Aharonov-Bohm string if there is
%a particle with a fractional charge which is charged under the remaining low-energy gauge group.
%Such a string can decay into flux. We will not study the process of a string decay into flux in this 
%paper.
In general, one can expect that this string is not 
charged under any scalar in the spectrum. 
This happens if, for example, every scalar charged under the anomalous $U(1)$
receives a vacuum expectation value.
Since it contains a magnetic flux tube, this string will 
break on monopoles. 

%%%%%%%%%%%%%%%%%%%%%%%%%%%%%%%%%%%%%%%%%%%%%%%%%%%%%%%%%%%%%%%%%%%%%%%%%%%%%%%%%%%%%%%%%%%%%%%%%%%%%%%%%%%

\subsection{Metastable Heterotic Strings}

%%%%%%%%%%%%%%%%%%%%%%%%%%%%%%%%%%%%%%%%%%%%%%%%%%%%%%%%%%%%%%%%%%%%%%%%%%%%%%%%%%%%%%%%%%%%%%%%%%%%%%%%%%%%

Despite the fact, that a heterotic cosmic string can break on monopoles, 
its lifetime can be sufficiently long.
The process of breakage of a string can be described by 
formation of a hole in the Euclidean worldsheet whose boundary
is the monopole worldline. The decay rate is then governed by the 
Euclidean action 
\begin{equation}
S_E[X^1, X^2] =m\int d\ell - \mu\int_{hole} dS.
\label{4.01}
\end{equation}
Here the first term is the length of the worldline, the second term 
is the area of the hole, $m$ is the monopole mass, $\mu$ is the string tension and 
$X^1$ and $X^2$ are the coordinates on the worldsheet.
A straightforward calculation shows that $S_E[X^1, X^2]$ is minimized by a circular 
worldline of the radius 
\begin{equation}
\frac{m}{\mu}.
\label{4.02}
\end{equation}
Substituting it to the action~\eqref{4.01}, we find that the decay rate behaves as
\begin{equation}
e^{-\pi m^2/2\mu}.
\label{4.15}
\end{equation}
It follows that,
if the square of the monopole mass is much bigger than the 
string tension, the lifetime of the string will be 
very long.
In this subsection, we will estimate the masses of heterotic monopoles
and show that they are indeed very large.

There are two sources of monopoles in heterotic M-theory. 
The first source is open membranes beginning on one of the orbifold planes,
ending on a five-brane and
wrapping a one-cycle of non-trivial $\pi_1$ in Calabi-Yau space. 
For these particle-like states to be magnetically charged, 
the five-brane should carry one or more $U(1)$ gauge groups. 
This means that this five-brane should be wrapped on a holomorphic 
curve of genus one or higher. To avoid the problem of breaking on these monopoles 
it is enough to require the absence of
such five-branes in the bulk.
Even if there are such five-branes, these monopoles will have very large masses 
comparing to the scale of the string tension as long as the five-branes are
far away from the orbifold fixed planes. This follows from a simple estimation.
The mass of this type of monopoles behaves as
\begin{equation}
m_1 \sim M_{11}^3 \ell_{CY} \ell,
\label{4.11}
\end{equation}
where $\ell_{CY}$ is the length of a one-cycle of non-trivial $\pi_1$ whose scale is set 
by the Calabi-Yau radius and $\ell$ is the distance between the five-brane and one of the 
orbifold planes. The scale of $\ell$ is set by the interval scale. It is easy to evaluate that 
for strings whose tension is within the bound~\eqref{2.2}, 
\begin{equation}
\frac{m_1}{\sqrt{\mu}} \sim 10^5.
\label{4.12}
\end{equation}
This implies that it will take a very long time for the strings to break. 

The second source is the traditional breaking of the GUT gauge group by Wilson lines.
The mass of monopoles of this type is of order 
\begin{equation}
m_2 \sim \frac{M_{GUT}}{g^2_{GUT}}.
\label{4.13}
\end{equation}
This means that
\begin{equation}
\frac{m_2^2}{\mu} \sim \frac{10}{g_{GUT}^2} \sim 4 \cdot 10^6.
\label{4.14}
\end{equation}
It follows from this equation that the decay rate~\eqref{4.15} is suppressed.
Thus, the strings discussed at the end of the previous subsection
have a good chance of being metastable with a sufficiently long lifetime. 

%%%%%%%%%%%%%%%%%%%%%%%%%%%%%%%%%%%%%%%%%%%%%%%%%%%%%%%%%%%%%%%%%%%%%%%%%%%%%%%%%%%%%%%%%%%%%%%%

\section{Acknowledgements}

%%%%%%%%%%%%%%%%%%%%%%%%%%%%%%%%%%%%%%%%%%%%%%%%%%%%%%%%%%%%%%%%%%%%%%%%%%%%%%%%%%%%%%%%%%%%%%%

The author is very grateful to Juan Maldacena and Joe Polchinksi
for helpful discussions and explanations. The work is 
supported by NSF grant PHY-0503584.

%%%%%%%%%%%%%%%%%%%%%%%%%%%%%%%%%%%%%%%%%%%%%%%%%%%%%%%%%%%%%%%%%%%%%%%%%%%%%%%%%%%%%%%%%%%%

%%%%%%%%%%%%%%%%%%%%%%%%%%%%%%%%%%%%%%%%%%%%%%%%%%%%%%%%%%%%%%%%%%%%%%%%%%%%%%%%%%%%%%%%%%%%%%%%%%%%%%%%%%%%%%

\end{document}